\documentstyle[12pt]{article}
\def\singlespace {\smallskipamount=3.75pt plus1pt minus1pt
                  \medskipamount=7.5pt plus2pt minus2pt
                  \bigskipamount=15pt plus4pt minus4pt
                  \normalbaselineskip=15pt plus0pt minus0pt
                  \normallineskip=1pt
                  \normallineskiplimit=0pt
                  \jot=3.75pt
                  {\def\smallskip {\vskip\smallskipamount}}
                  {\def\medskip   {\vskip\medskipamount}}
                  {\def\bigskip   {\vskip\bigskipamount}}
                  {\setbox\strutbox=\hbox{\vrule
                    height10.5pt depth4.5pt width 0pt}}
                  \parskip 7.5pt
                  \normalbaselines}
\def\middlespace {\smallskipamount=5.625pt plus1.5pt minus1.5pt
                  \medskipamount=11.25pt plus3pt minus3pt
                  \bigskipamount=22.5pt plus6pt minus6pt
                  \normalbaselineskip=22.5pt plus0pt minus0pt
                  \normallineskip=1pt
                  \normallineskiplimit=0pt
                  \jot=5.625pt
                  {\def\smallskip {\vskip\smallskipamount}}
                  {\def\medskip   {\vskip\medskipamount}}
                  {\def\bigskip   {\vskip\bigskipamount}}
                  {\setbox\strutbox=\hbox{\vrule
                    height15.75pt depth6.75pt width 0pt}}
                  \parskip 11.25pt
                  \normalbaselines}
\def\doublespace {\smallskipamount=7.5pt plus2pt minus2pt
                  \medskipamount=15pt plus4pt minus4pt
                  \bigskipamount=30pt plus8pt minus8pt
                  \normalbaselineskip=30pt plus0pt minus0pt
                  \normallineskip=2pt
                  \normallineskiplimit=0pt
                  \jot=7.5pt
                  {\def\smallskip {\vskip\smallskipamount}}
                  {\def\medskip   {\vskip\medskipamount}}
                  {\def\bigskip   {\vskip\bigskipamount}}
                  {\setbox\strutbox=\hbox{\vrule
                    height21.0pt depth9.0pt width 0pt}}
                  \parskip 15.0pt
                  \normalbaselines}
\def\be{\begin{equation}}
\def\ee{\end{equation}}
\def\bearr{\begin{eqnarray}}
\def\bearrs{\begin{eqnarray*}}
\def\eearr{\end{eqnarray}}
\def\eearrs{\end{eqnarray*}}
\def\barr{\begin{array}}
\def\earr{\end{array}}

\def\lessapprox{\stackrel{_<}{_\sim} }
\newcommand{\Redtau}{{\rm Re}\ d_\tau}
\newcommand{\Imdtau}{{\rm Im}\ d_\tau}
\begin{document}
\thispagestyle{empty}

\begin{flushright}
PRL-TH-94/32\\ UNIL-TP-5/94\\ hep-ph/9411399
\end{flushright}

\middlespace
\begin{center}
{\Large\bf Measurement of the $\tau$ electric dipole moment using
longitudinal polarization of $e^+e^-$ beams}
\vspace{0.1in}

{\large B. Ananthanarayan}\\
{\bf Institut de Physique Th\'eorique,
Universit\'e de Lausanne\\ CH 1015 Lausanne, Switzerland}
\vspace{.1in}
{\large Saurabh D. Rindani}\\
{\bf Theory Group, Physical Research Laboratory\\
Navrangpura, Ahmedabad\ 380 009, India}
\end{center}

\begin{abstract}
Certain CP-odd momentum correlations in the production and
subsequent decay of $\tau$ pairs in $e^+ e^-$ collisions are
enhanced significantly when the $e^+$ and $e^-$
beams are longitudinally polarized. These may be used to probe the real and
imaginary parts of $d_\tau^\gamma$, the electric dipole moment of the $\tau$.
Closed-form expressions for these ``vector correlations'' and
the standard deviation of the operators defining them
due to standard model interactions are
presented for the two-body final states of $\tau$ decays.
If 42\% average polarization of each beam
is achieved, as proposed for the tau-charm factories,
with
equal integrated luminosities for each sign of polarization
and a total yield of
$2\cdot 10^7$ $\tau^+ \tau^-$ pairs,
it is
possible to attain sensitivities for $\vert\delta {\rm
Re} d_{\tau}^{\gamma}\vert$ of $8\cdot 10^{-19}$, $1\cdot
10^{-19}$, $1\cdot 10^{-19}$
$e$ cm respectively and for $\vert\delta {\rm Im}
d_{\tau}^{\gamma}\vert$ of $4\cdot 10^{-14}$,
$6\cdot 10^{-15}$, $5\cdot 10^{-16}$ $e$ cm respectively at the
three operating center-of-mass energies of 3.67, 4.25 and 10.58
GeV.   These bounds emerge when the effects of a posible weak dipole
form factor $d_\tau^Z$ are negligible as is the case when it is of
the same order of magnitude as $d_\tau^\gamma$.  Furthermore, in
such a polarization experiment where different polarizations are
possible, a model-independent disentangling of their individual
effects is possible, and a technique to achieve this is described.
A strong
longitudinal polarization physics programme at the tau-charm factory
appears warranted.
\end{abstract}
\newpage
\noindent {\large\bf I. Introduction}

Leptonic CP violation would signal interactions not described in
the framework of the standard model since it arises there only at the
multi-loop level and is way below any measurable level [1]. The
presence of a non-zero and large electric dipole moment (edm) of
any elementary particle is a signature of CP-violating
interactions [2]. Whereas the edm of the electron is
constrained to be $\lessapprox 10^{-26}$ $e$ cm and that of the muon
is $\lessapprox 10^{-19}$ $e$ cm [3], the constraint on the edm of
the $\tau$ lepton [4] is less stringent, viz., $\lessapprox
5\cdot 10^{-17}$ $e$ cm [3]. Thus an important experimental challenge is
to measure the $\tau$ electric dipole moment far more
accurately than at present. The analogous coupling of
the $\tau$ to the $Z$ boson, the weak dipole form factor (wdff), is
better constrained from LEP data be $\lessapprox 3.7 \cdot 10^{-17}$
$e$ cm at the $Z$ resonance [4]. It has recently been proposed [5] that the
availability of large polarization at SLC might improve this
measurement some more. The purpose of this note is to
demonstrate that the availability of large polarization will go
a long way in improving the measurement of the $\tau$ edm.
Further, the discussion presented here may also be easily
extended to other physiucal situations which include the
measurement of CP-violating form factors in $W^+ W^-$ or
$t\bar{t}$ production.

The approach proposed  consists
measuring CP-odd correlations [6,7] amongst the momenta of the
final state particles in the reaction $e^+ e^- \rightarrow \tau^+ \tau^-
\rightarrow X^+ \bar{\nu}_{\tau} X^- \nu_{\tau}$. In particular, one
may construct scalar,
vector and tensor correlations [8] from the momenta ${\bf q}_+$ and
${\bf q}_-$ of the decay products of the $\tau^+$ and $\tau^-$.
One such tensor has been used to constrain the real part of
the $\tau$ wdff from LEP data [9] where the $Z$ contribution dominates the
cross-section. Indeed, other tensor correlations have been found
to be sensitive to the imaginary part of the wdff as well [10]
and may be used at LEP to constrain it in the event of the
absence of a significant non-zero measurement of such
correlations. In [5] it has been shown that the presence of large
longitudinal polarization renders certain simple vector correlations sensitive
to the real as well as to the imaginary parts of the wdff at the
$Z$ factory SLC.

Here we investigate the sensitivity of these
correlations to the real and imaginary parts of the edm  when
the  production of $\tau^+ \tau^-$ is no longer dominated by $Z$
exchange and instead by photon  exchange as is typically
the case when $\sqrt{s} \ll m_z$. In particular, we will present
much of our numerical results for the proposed tau-charm
factories ($\tau$cF) [11] where there exists an ample opportunity to have
substantial polarization of the $e^+$ and $e^-$ beams [12]. The
prospects for  the measurement of the edm at the tau-charm
factory with unpolarized beams has already been considered [10]
by measuring tensor correlations amongst the momenta of final
state particles in the $\tau$ decays.
Algebraically our approach proves simpler since
the vector correlations (more correctly their scalar product
with the $e^+$ beam direction) we consider can be expressed in
closed form and the standard deviation of the operators
defining the correlations due to the standard
model interactions can also be so expressed for the two-body
final states of the $\tau$ decays. In practice the expressions of
Ref.[5] valid at the $Z$ peak are now generalized to include the
pure $\gamma^*$ as well as the $\gamma^*-Z^*$ interference
terms, using in addition to SM,
the CP-violating terms in the effective Hamiltonian for the reaction
\be
e^+ e^- \rightarrow \gamma^*,Z^* \rightarrow \tau^+ \tau^-
\rightarrow \overline{B} A \overline{\nu}_{\tau} \nu_{\tau}
\ee
given in Ref.[10].
(Note that the expressions obtained
here are also valid at much larger center of mass
energies where contributions from $\gamma^*$ and $Z^*$ are
significant, and can also be easily modified for $W^+ W^-$ and
$t \bar{t}$ production [13]
where it may be possible to probe CP violation).
For comparable
magnitudes of the edm and wdff, at the $\tau c$F energies,
the CP-odd correlations obtain their most significant
contribution from the edm [14].

 The CP-odd momentum
correlations we consider here are associated with the c.m. momenta
${\bf p}$ of $e^+$,
${\bf q}_{\overline{B}}$ of $\overline{B}$ and
${\bf q}_A$ of $A$, where the $\overline{B}$ and
$A$ arise in the decays $\tau^+ \rightarrow \overline{B}+
\overline{\nu}_{\tau}$
and $\tau^-\rightarrow A + \nu_{\tau}$, and where $A,\ B$ run over
$\pi$, $\rho$, $A_1$, etc.  In the case when $A$ and $B$ are
different, one has to consider also the decays with $A$ and $B$
interchanged, so as to construct correlations which are
explicitly CP-odd. The calculations
include
two-body decay modes of the $\tau$ in general
and is applied specifically to the case of $\tau\rightarrow
\pi+\nu_\tau$ and $\tau\rightarrow \rho+\nu_\tau$ due to the
fact that these modes possess a good resolving power
of the $\tau$ polarization, parametrized in terms of
the constant $\alpha$ which takes the value 1 for the
$\pi$ channel (with branching fraction of about
$11\%$) and 0.46 for the $\rho$ channel [9] (with branching
fraction of about $22\%$).
It may be noted that with these final states the
substantive fraction of the channels that are sensitive
to such correlations are accounted for; three-body
leptonic final states must also be included; they
are characterized by a somewhat smaller $\alpha=-0.33$
(with branching fraction of
about $35\%$).  Thus with the channels studied here, one
more or less reaches the limits of discovery in
such experiments.
(It would
also be possible to apply this to the decay $\tau\rightarrow
A_1+\nu$;  $\alpha_{A_1}$ is however too small
to be of any experimental relevance.)
Further, we
also present closed-form expressions for the variance
of the correlations considered due to standard model
interactions.
 These, because of finite statistics, provide a
measure of the CP-invariant background to the determination of
the CP-odd contributions to the correlations. In case of a
negative result, the limit on the CP-violating interactions is
obtained using the value of the variance and the size of the
data sample.

It must be noted that correlations which are CP violating in the
absence of initial beam polarization are not strictly CP odd
for arbitrary $e^+$ and $e^-$ polarizations, since the initial
state is then not necessarily  CP even. We argue, however,
that this is true to a high degree of accuracy in the case at
hand. Besides, for our numerical results, we restrict ourselves
to the case where the $e^+$ and $e^-$ polarizations are equal
and opposite, thus making up a CP-even initial state.

We follow a slightly different
notation notation from Bernreuther {\it et al.} [10] and use the
symbols $B_i$ and $B_j$ to denote the intermediate
vector bosons, the photon and the Z.  In the mean as well
as in the variances and in the cross-sections the contributions
would eventually
have to be summed over $i,\ j$.
Our main result is that the contribution to
 certain CP-odd correlations, which are
relatively small in the absence of polarization, since they
come with a factor $r_{ij}=(V_e^iA_e^j+V_e^jA_e^i)/(V_e^iV_e^j+A_e^iA_e^j)$
and get enhanced in the presence of
polarization, now being proportional to
$(r_{ij} - P)$, with the corresponding contribution
to the cross-section being multiplied by $(1-r_{ij}P)$.
Here $V_e^i, A_e^i$ are the
vector and axial vector couplings of $e^-$ to $B_i$, and $P$ is
the effective polarization  defined by \[
P=\frac{P_e-P_{\overline{e}}}{1-P_eP_{\overline{e}}}, \]
where $P_e\ (P_{\overline{e}})$ is the polarization of
the electron (positron) and is positive for right-circular
polarization for each particle in our convention.

The
correlations which have this property are those which have an
odd number of factors of the $e^+$ c.m. momentum
${\bf p}$, since this would need P and C
violation at the electron vertex.  Furthermore, we suggest a
procedure for  obtaining these correlations from the difference in the
event distributions for a certain polarization $P$ and the
sign-flipped polarization $-P$. With this procedure, the correlations
are further enhanced, leading
to increased sensitivity.
The inclusion of the $\rho$ channel leads to a considerable
improvement in the sensitivity that can be reached in the
measurement of Im $d_{\tau}$ while improving the
measurement of Re $d_\tau$ less spectacularly.

More specifically, we have considered the observables $O_1
\equiv \frac{1}{2} \left[\hat{\bf p}\cdot \left({\bf q}_{\overline B}\times
{\bf q}_A \right)\right.$\linebreak $+\left.\hat{\bf p}\cdot
\left({\bf q}_{\overline A}\times  {\bf q}_B \right)\right]$ and $O_2 \equiv
\frac{1}{2}\left[\hat{\bf p}\cdot \left({\bf q}_A + {\bf
q}_{\overline B} \right)+\hat{\bf p}\cdot \left({\bf q}_{\overline A} +
{\bf q}_B \right)\right]$ (the caret denoting a unit vector) and
obtained analytic
expressions for their mean values and standard deviations in the
presence of longitudinal polarization.  $O_2$,
being CPT-odd, measures ${\rm Im}\,d_{\tau}^i$, whereas $O_1$
measures ${\rm Re}\,d_{\tau}^i$.
Inclusion of other exclusive $\tau$
decay modes (not studied here) would improve the sensitivity
further.

As a result, we find it possible to define 1 s.d. sensitivities
$\mid \delta{\rm Re} d^{\gamma}_\tau\mid$ and $\mid \delta{\rm Im}
 d^{\gamma}_\tau\mid$
from the two-body decay modes when we make the
reasonable assumption that the edm and wdff are of comparable
magnitudes. To facilitate comparison with
Ref.[10] we assume center of mass energies of 3.67, 4.25 and
10.58 GeV.

In order to answer what makes our correlations viable, we now
discuss what prospects exist for longitudinal polarization at
the $\tau c$F [15].
One proposal [12] is that the $e^+$ and $e^-$ beams be polarized
in separate rings to achieve an average degree of polarization
of each beam as large as 42\% before being injected in tho the
main ring. (It is also important to note that this would not
lead to a large loss in luminosity, in contrast to the situation
at linear colliders where the large polarization is accompanied
by modest luminosities as, for instance, in the case of SLC).
This proposal also envisages all four possiblities in the
combinations of the polarizations. In particlular, as an
effective polarization $P$
can be as large as 0.71 and of either sign in the $e^+e^-$
collisions at the $\tau$cF. We show that with
equal integrated luminosities with either sign,
$\int {\cal L}(P)dt=\int {\cal L}(-P) dt$ and
a total yield
$N_{\tau^+\tau^-}$
of $2\cdot 10^7$ $\tau^+\tau^-$ pairs,
we can
probe
the real part of the edm of the $\tau$ to the remarkable 1
s.d. precision of $\sim 10^{-19}~e$ cm. The imaginary
part however is not probed to such a spectacular degree.
We finally describe a technique whereby the reasonable
assumption of the comparability of magnitudes of the
edm and wdff can be avoided in such a polarization experiment.
These
considerations enable us to build a very strong case for
introducing longitudinal polarization at the $\tau$cF [16, 17].

\bigskip

\noindent{\large\bf II. Notation and Formalism}

Although much of this section has already been described
in our previous papers [5] we will repeat it for the sake
of completeness and to make the generalization to the inclusion
of $\gamma$ and $Z$ (we drop the asterisk in what follows
since no confusion is bound to arise) more transparent.

The process we consider is
\begin{equation}
e^-(p_-) + e^+(p_+)\rightarrow \tau^-(k_-) + \tau^+(k_+),
\end{equation}
with the subsequent decays
\begin{equation}
\tau^-(k_-)\rightarrow A(q_A) + \nu_{\tau} ,\;
\tau^+(k_+)\rightarrow \overline{B}(q_{\overline B}) +\overline{\nu}_{\tau} ,
\end{equation}
together with decays corresponding to $A$ and $B$ interchanged
in (2).

Under CP, the various three-momenta transform as
\begin{equation}
{\bf p}_-\leftrightarrow -
{\bf p}_+ ,\;
{\bf k}_-\leftrightarrow -
{\bf k}_+ ,\;  {\bf q}_{A,B}
\leftrightarrow - {\bf q}_{{\overline A},{\overline B}} .
\end{equation}
We choose for our analysis the two CP-odd observables $O_1
\equiv \frac{1}{2} \left[\hat{\bf p}\cdot \left({\bf q}_{\overline B}\times
{\bf q}_A \right)\right.$\linebreak $+\left.\hat{\bf p}\cdot
\left({\bf q}_{\overline A}\times  {\bf q}_B \right)\right]$ and $O_2 \equiv
\frac{1}{2}\left[\hat{\bf p}\cdot \left({\bf q}_A + {\bf
q}_{\overline B} \right)+\hat{\bf p}\cdot \left({\bf q}_{\overline A} +
{\bf q}_B \right)\right]$, which have an odd number of factors
of $\hat{\bf p}$, the unit vector along ${\bf p}_+$.  As
mentioned before, they are expected to get
enhanced in the presence of polarization.

Though these observables are CP
odd, their observation with polarized $e^+$ and $e^-$ beams is
not necessarily an indication of CP violation, unless the $e^+$
and $e^-$ longitudinal polarizations are equal and opposite, so
that the initial state is described by a CP-even density matrix.
The case when only the $e^-$ is
polarized, has already been discussed [5]. Though our
expressions for correlations will be valid for arbitrary
polarizations, our results will be only for equal and opposite
electron and positron polarizations, so that the correlations
are strictly CP odd.

Of $O_1$ and  $O_2$, $O_1$ is even under the combined CPT transformation,
and $O_2$ is CPT-odd. A CPT-odd
observable can only have a non-zero value in the presence of an
absorptive part of the amplitude.  It is therefore expected that
$\langle O_2\rangle $ will be proportional to the imaginary part of the
dipole form factors ${\rm Im}\,d_{\tau}^i$ , since final-state
interaction, which could give rise to an absorptive part,
is negligible in the weak $\tau$ decays. Since $\langle O_1\rangle $ and mean
values of other CPT-even quantities will be proportional to
${\rm Re}\,d_{\tau}^i$, phase information on
$d_{\tau}^i$ can only be obtained if $\langle O_2\rangle $ (or
some other CPT-odd quantity) is also measured.

We assume SM couplings for all particles except $\tau$, for which
an additional edm and wdff interaction is assumed, viz.,
\begin{equation}
{\cal L}_{CPV} =
- \frac{i}{2} d_{\tau}^Z
\overline{\tau}\sigma^{\mu\nu}\gamma_5\tau
\left(\partial_{\mu}Z_{\nu} - \partial_{\nu}Z_\mu\right)
- \frac{i}{2} d_{\tau}^\gamma
\overline{\tau}\sigma^{\mu\nu}\gamma_5\tau
\left(\partial_{\mu}A_{\nu} - \partial_{\nu}A_\mu\right) ,
\end{equation}
Using (5), we now proceed to calculate
$\langle O_1\rangle $ and $\langle O_2\rangle $ in the presence
of an effective longitudinal polarization $P$.

We can anticipate the effect of $P$ in general for the process
(1).  We can write the matrix element squared for the process in
the leading order in perturbation theory, neglecting the
electron mass, as
\begin{equation}
\vert M\vert^2 = \sum_{i,j} L^{ij}_{\mu\nu}(e) L^{ij
\mu\nu*}(\tau) \frac{1}{s-M^2_i}\, \frac{1}{s-M^2_j} ,
\end{equation}
where the summation is over the gauge bosons $(\gamma ,Z,\ldots)$
exchanged in the $s$ channel, and $L^{ij}_{\mu\nu}(e,\tau)$
represent the tensors arising at the $e$ and $\tau$ vertices:
\begin{equation}
L^{ij}_{\mu\nu} = V^i_\mu V^{j*}_{\nu} .
\end{equation}
For the electron vertex, with only the SM  vector and axial-vector
couplings,
\begin{equation}
V^i_{\mu}(e) = \overline{v}(p_+ ,
s_+)\gamma_{\mu}\left(V_e^i-\gamma_5A_e^i\right)u(p_- ,
s_-),
\end{equation}
We
have the definitions
\begin{equation}
V_{e/\tau}^{\gamma} = - e,\; A_{e/\tau}^{\gamma} = 0 ;
\end{equation}
\begin{equation}
V_{e/\tau}^Z = (-\frac{1}{2} + 2\sin ^2\theta_W)\frac{e}{2\sin\theta_w
\cos\theta_w} ,\; A_{e/\tau}^Z =
(-\frac{1}{2})\frac{e}{2\sin\theta_w\cos\theta_w} .
\end{equation}
It is easy to check, by putting in helicity projection operators,
that
\begin{eqnarray}
\lefteqn {L^{ij}_{\mu\nu}(e)= } && \nonumber \\
 & \! &\!\!\!\! \left\{\left[\left( 1-P_eP_{\overline{
e}}\right) \left(V_e^iV_e^j
+ A_e^iA_e^j\right) - \left( P_e-P_{\overline
e}\right)\left(V_e^iA_e^j +
A_e^iV_e^j\right) \right]
Tr(\rlap{$p$}/_-\gamma_{\mu}\rlap{$p$}/_+\gamma_\nu) \right.
\nonumber \\
 &\! +&\!\!\!\! \left.  \left[ \left( P_e-P_{\overline e}\right)
\left(V_e^iV_e^j + A_e^iA_e^j\right) -
\left( 1-P_eP_{\overline e}\right)\left(V_e^iA_e^j +
A_e^iV_e^j\right)\right]
Tr \left(\gamma_5\rlap{$p$}/_-
\gamma_{\mu}\rlap{$p$}/_+\gamma_\nu \right) \right\} \nonumber\\
&&
\end{eqnarray}
in the limit of vanishing electron mass, where $P_e$ ($P_{\overline
e}$) is the
degree of the $e^-$ ($e^+$) longitudinal polarization.
Eq.(11) gives a simple way of incorporating the effect of the
longitudinal polarization.
\begin{equation}
\begin{array}{clclc}
V_e^iV_e^j+ A_e^iA_e^j &\rightarrow &V_e^iV_e^j+A_e^iA_e^j &-& P \,
\left( A_e^iV_e^j+A_e^jV_e^i \right),  \\
(A_e^iV_e^j+A_e^jV_e^i) &\rightarrow &(A_e^iV_e^j+A_e^jV_e^i) &-
&P \left( V_e^iV_e^j+A_e^iA_e^j \right),
\end{array}
\end{equation}
where $P$ is as defined earlier.

To calculate correlations of $O_1$ and $O_2$, we need the
differential cross section for (1) followed by (2)
arising from SM $\gamma$ and $Z$ couplings of $e$ and $\tau$, together
dipole couplings of $\tau$ arising from eq.(4). The calculation may be
conveniently done, following ref.[10], in
steps, by first determining the production matrix $\chi$ for
$\tau^+\tau^-$ in spin space, and then taking
its trace with the decay matrices $\cal{D}^{\pm}$ for $\tau^{\pm}$
decays into single charged particle in addition to the
invisible neutrino.

The differential cross section for (1) is  given by
\begin{equation}
\frac{d\sigma}{d\Omega_kd\Omega^*_-d\Omega^*_+dE^*_-dE^*_+} =
\frac{k}{8\pi s}\frac{1}{(4\pi)^3}
\chi^{\beta\beta^{\prime}, \alpha\alpha^{\prime}}
{\cal D}^-_{\alpha^\prime\alpha} {\cal D}^+_{\beta^\prime\beta},
\end{equation}
where $d\Omega_k$ is the solid angle element for
${\bf k}_+$ in the overall c.m. frame, $k =
\vert{\bf k}_+\vert$, and $d\Omega^*_{\pm}$ are
the solid angle elements for ${\bf q}^*_{{\overline B},A}$,
the $\overline{B}$ and $A$
momenta in the $\tau^{\pm}$ rest frame.  The
${\cal D}$ matrices are given by
\begin{eqnarray}
& {\cal D}^{+} = \delta\left(E^*_{B} - E_{0B}\right)\left[ 1 -
\alpha_B {\bf \sigma}_{+} \cdot\hat{\bf q}^*_{B} \right] & \nonumber \\
& {\cal D}^{-} = \delta\left(E^*_{A} - E_{0A}\right)\left[ 1 +
\alpha_A {\bf \sigma}_{-} \cdot\hat{\bf q}^*_{A} \right], &
\end{eqnarray}
where ${\bf \sigma}_{\pm}$ are the Pauli
matrices corresponding to the $\tau^{\pm}$ spin, $E^*_{\pm}$ are
the charged particle energies in the $\tau^{\pm}$ rest frame, and
\begin{equation}
E_{0A,B} = \frac{1}{2} m_{\tau} (1 + p_{A,B}) ;\;
p_{A,B} = m^2_{A,B}/m^2_{\tau}.
\end{equation}

The expressions for $\chi$ arising from SM as well as the
CP-violating form factor
couplings of $\tau$ are rather long, and we refer the reader to
ref.[10] for these expressions in the absence of polarization.
It is straightforward to incorporate polarization using (12).

\bigskip

\noindent{\large {\bf III. Results}}

Using eqns. (13)-(15) above, as well as the expression for the
$\tau^+\tau^-$ production matrix $\chi$ from [10], we can obtain
expressions for $\langle O_1 \rangle$ and $\langle O_2\rangle$
by writing $O_1$ and $O_2$ in terms of the $\tau$ rest frame
variables and performing the integrals over them analytically. The
expressions for the correlations $\langle O_1\rangle $ and
$\langle O_2\rangle$ obtained
are, neglecting $\sum_{i,j}d^i_{\tau}d^j_{\tau}$,
\begin{eqnarray} \nonumber
& \langle O_1\rangle   =  -
{{1}\over{36x\sigma}}
\sum_{i,j}K_{ij}s^{3/2}m_{\tau}^2
(1-x^2)\left(\frac{r_{ij} - P}{1-r_{ij}P}\right) &  \nonumber \\
 & [(A_\tau^i\Redtau^j+A_\tau^j\Redtau^i)
\alpha_A \alpha_B(1-p_A)(1-p_B) - & \nonumber \\
& \frac{3}{2}(V_\tau^i\Redtau^j+V_\tau^j\Redtau^i)
[\alpha_A(1-p_A)(1+p_B)+
\alpha_B(1-p_B)(1+p_A)], &
\end{eqnarray}
and
\begin{eqnarray}
& \langle O_2\rangle = {{1}\over{3\sigma}}
\sum_{i,j}K_{ij}s^{3/2}m_{\tau}
\left(\frac{r_{ij} - P}{1-r_{ij}P}\right) & \nonumber \\
& \frac{1}{4}(A_\tau^{i}\Imdtau^j+A_\tau^j\Imdtau^{i})(1-x^2)
(\alpha_A(1-p_A)
+\alpha_B(1-p_B)),&
\end{eqnarray}
where $x =
2m_{\tau}/\sqrt{s}$ and $\sigma$, which is  the
cross-section apart from a normalization factor, is given by:
\begin{equation}
\sigma=\sum_{i,j}K_{ij}s[V_\tau^iV_\tau^j(1+\frac{x^2}{2})+
A_\tau^iA_\tau^j(1-x^2)],
\end{equation}
\noindent and
\begin{equation}
K_{ij}=\frac{s(V_e^iV_e^j+A_e^iA_e^j)(1-r_{ij}P)}
{(s-M_i^2)(s-M_j^2)}.
\end{equation}

Here we neglect the width of the $Z$ since we work now at
$\sqrt{s}<<m_Z$. However at the $Z$ peak we neglect $\gamma$ and
treat the system in the narrow-width approximation.

We have also obtained analytic expressions for the variance
$\langle  O^2\rangle  - \langle  O\rangle ^2 \approx \langle  O^2\rangle $
in each case, arising from the CP-invariant SM part of the
interaction:
\begin{eqnarray}\nonumber
  &   \langle O_1^2\rangle  =
{{1}\over{720x^2\sigma}}\sum_{i,j}K_{ij}s m_\tau^4
 \biggl(
(1-p_A)^2(1-p_B)^2 & \nonumber \\
&
  [V_\tau^iV_\tau^j(6+8x^2+x^4)+
A_\tau^iA_\tau^j(6-2x^2-4x^4)] &  \nonumber \\
  & +(1-x^2)
\left([(1+p_A)^2(1-p_B)^2+(1+p_B)^2(1-p_A)^2] \right. & \nonumber \\
& \left. [3V_\tau^iV_\tau^j(3+2x^2)+9A_\tau^iA_\tau^j(1-x^2)
] \right. & \nonumber \\
& \left. +4\alpha_A\alpha_B
(1-p_B^2)(1-p_A^2)(1-x^2)[V_\tau^iV_\tau^j
-A_\tau^iA_\tau^j]\right)   & \nonumber \\
  & -6(1-p_A)(1-p_B)(V_\tau^iA_\tau^j+V_\tau^jA_\tau^i)
(1-x^2)(1-\frac
{x^2}{6}) & \nonumber \\
 & [\alpha_A (1+p_A)(1-p_B)+\alpha_B (1+p_B)(1-p_A)]
\biggr), &
\end{eqnarray}
\begin{eqnarray}
  & \langle O_2^2\rangle =
{{1}\over{360x^2\sigma}}\sum_{i,j}K_{ij}s m_\tau^2 & \nonumber \\
  & \biggl[ \biggl( 3[(1-p_A)^2+(1-p_B)^2][V_\tau^i V_\tau^j
(4+7x^2+4x^4)+A_\tau^iA_\tau^j 2(1-x^2)(2+3x^2)]
 & \nonumber \\
&  -2\alpha_A \alpha_B (1-p_A) (1-p_B)
[V_\tau^iV_\tau^j(4+7x^2+4x^4)+A_\tau^iA_\tau^j
4(1-x^2)^2]\biggr)
  & \nonumber \\
  & +6
\biggl( 6(1-x^2)(p_A-p_B)^2[V_\tau^iV_\tau^j(1+\frac{x^2}{4})
+A_\tau^iA_\tau^j(1-x^2)] &  \nonumber \\
  & -(V_\tau^iA_\tau^j+V_\tau^jA_\tau^i)(1-x^2)(4+x^2)(p_A-p_B)
[\alpha_A(1-p_A)-\alpha_B(1-p_B)]\biggr)\biggr]. &
\end{eqnarray}

The results for the significant two-body
decay channels are presented in the tables.  In Tables 1-6
 we have presented, for three typical values of
$\sqrt{s}$ at which the $\tau$cF is expect to run,
  the values of
$c_{AB}$ for $O_1$ and $O_2$ respectively, defined as
 the correlation for a value of ${\rm Re}\,d_{\tau}^\gamma$ or
${\rm Im}\,d_{\tau}^\gamma$ (as the case may be) equal to
$e/\sqrt{s}$, for some values of $P$ chosen to correspond to
average beam polarizations of 0, 35\%, 42\% and 100\%.
We have also presented the value of $\sqrt{\langle O^2_a\rangle },\
(a=1,2)$.
 This 1 s.d. limit is the value
of $d_{\tau}^\gamma$ which gives a mean value of $O_a$
equal to the s.d. $\sqrt{\langle O^2_a\rangle /N_{AB}}$ in each case:
\begin{equation}
c_{AB}^{1(2)}\delta {\rm Re(Im)}
d_\tau^\gamma=\frac{e}{\sqrt{s}}\frac{1}{\sqrt{N_{AB}}}
\sqrt{\langle O^2_{1(2)}\rangle}.
\end{equation}
Here $N_{AB}$ is the number of events in the channel $A\overline{B}$
(or $\overline{A}B$), and is given by
\begin{equation}
N_{AB}=N_{\tau^+\tau^-}B(\tau^-\rightarrow
A\nu_\tau )B(\tau^+\rightarrow \overline{B}\overline{\nu}_\tau),
\end{equation}
where we take $N_{\tau^+\tau^-}(P)=10^7$.

These limits can be improved by looking at
correlations of the same observables, but in a sample obtained
by counting the difference between the number of events for a certain
polarization, and for the corresponding sign-flipped polarization.
If the partial cross section for the process for a
polarization $P$ is given by
\begin{equation}
d\sigma(P)=\sum_{i,j}\left\{ (X_{ij}+r_{ij}Y_{ij})-P(r_{ij}X_{ij}
+ Y_{ij}) \right\},
\end{equation}
we can define a polarization asymmetrized distribution
\begin{equation}
\vert d\sigma(P) - d\sigma(-P)\vert = 2  \vert P \sum_{i,j} (r_{ij}X_{ij}
+ Y_{ij})\vert.
\end{equation}
We can then compute the mean and standard deviation for the
correlations over this distribution and these are tabulated in
Tables 7-9. The correlations get contributions from the $\pm 2P
\sum Y_{ij}$ term in eq.(25) as compared to the
$\sum r_{ij}Y_{ij}$ and is therefore enhanced, since $\vert
r_{ij}\vert <1$.
However the sensitivities are now
computed for smaller event
samples whose size is given by $\vert P\sum_{i,j} r_{ij}N_{ij}\vert$ where
$\sum_{i,j} N_{ij}$ stands for the total number of $\tau^+\tau^-$
pairs including both polarizations $P$ and $-P$.
The standard deviations are only slightly affected. The net
result is an increase in the sensitivity.
For the different
values of $\sqrt{s}$ we tabulate  the associated quantity,
${{\vert \sum_{i,j}r_{ij}N_{ij}\vert }\over{
\sum_{i,j}N_{ij}}}$, the effective polarization asymmetry in Table 10.
Indeed, the improvement in sensitivity is seen to be by an order
of magnitude.

We can combine the sensitivities from the different $\tau$
channels in inverse quadrature, to get the improved numbers for
$\vert\delta {\rm
Re} d_{\tau}^{\gamma}\vert$ of $8\cdot 10^{-19}$, $1\cdot
10^{-19}$, $1\cdot 10^{-19}$
$e$ cm respectively and for $\vert\delta {\rm Im}
d_{\tau}^{\gamma}\vert$ of $4\cdot 10^{-14}$,
$6\cdot 10^{-15}$, $5\cdot 10^{-16}$ $e$ cm respectively at the
three  center of mass energies of 3.67, 4.25 and 10.58
GeV.

Thus far and for the purposes of Tables 1-9,
we have made the altogether reasonable assumption that
the contribution of the wdff is negligible which is justified
so long as the edm and wdff are of comparable magnitude.
However, no such assumption is really necessary in polarization
experiments such as these where the ability to run the experiment
at different polarizations allows one to disentangle their
individual contributions to the correlations considered here.
Indeed, it has been pointed out in the context of CP violation
in the $t\bar{t}$ system that varying the polarization allows a
model independent determination of the separate contributions
of the edm and wdff to the correlations of the type considered
here [13].  The principle is that at a given polarization, a certain
linear combination of the two form factors alone can be measured.
Performing the experiment at two different polarizations enables
us to disentangle the two form factors.  Similarly, the 1 s.d.
limits also can only be placed on such a linear combination.
Indeed, such 1 s.d. limits would be defined by straight lines
given by equations such as
\begin{equation}
\delta{\rm Re}d_\tau^\gamma/a+\delta{\rm Re}d_\tau^Z/b = \pm 1
\end{equation}
\noindent for the limits arising from $O_1$ and by
\begin{equation}
\delta{\rm Im}d_\tau^\gamma/c+\delta{\rm Im}d_\tau^Z/d = \pm 1
\end{equation}
\noindent for the limits arising from $O_2$
where the numbers $a,\ b,\ c$ and $d$ can be explicitly
computed for a given $P$ and $N$. This is also presented for the
polarization asymmetrized distribution for which we have set $P=1$
(with the understanding that this would have to be
scaled by $\sqrt{P}$ if $P$ is the polarization realized in
a certain experiment).
In particular, $a$ ($c$) is the sensitivity
of the  real (imaginary) part of the edm in inverse quadrature
when the real (imaginary) part of the wdff is set to zero and
$b$ ($d$) is the sensitivity of the
real (imaginary) part of the wdff in inverse quadrature
when the real (imaginary) part of the edm is set to zero.
We tabulate these quantities
for the three different c.m. energies and for different
polarizations for the parent distributions in Table 11 and
for the asymmetrized distributions in Table 12.  Note that
one can read from the columns for $a$ and $c$ in Table 12
the bounds cited in the abstract and in the
preceding paragraph (scaled by
$\sqrt{0.71}$). These implicitly assumes
that the magnitudes of the edm and wdff are comparable and
therefore the latter may be ignored for such considerations.

We now discuss how the results of Table 11 may be used in order
to essentially define regions in the Re (Im) $d_\tau^\gamma$--
$d_\tau^Z$ planes due to finite statistics, say at the
1 s. d. level.  By performing the experiment at two values
of $P$, say $P_1$ and $P_2$, one obtains two sets of straight
lines defined above.  The vertices of the interesection of these
4 lines defines the parallellogram in each of these planes which
cannot be ruled out due to the finite statistics.  The best results
may be obtained by taking the largest value of polarization realizable
$P_{max}$ and taking $P_1=-P_2=P_{max}$.  In Table 13 we tabulate for the
three different values of $P_{max}$ two pairs $(A,B)$ for the
Real and $(C,D)$ for the Imaginary
planes, which give the coordinates of two vertices in the
Real and Imaginary $d_\tau^\gamma$--
$d_\tau^Z$ planes respectively with $(-A,-B)$ and
$(-C,-D)$ giving the remaining pairs. Thus the availability
of polarization and of either sign provides for a model independent
scheme for constraining regions of the parameter space spanned by
the CP-violating form factors.  It must be noted that the price to be
paid for such a model-independent bound on each of the form factors
is large.  In particular, from Table 13,
the most stringent such bound on the
magnitude of
${\rm Re}({\rm Im})d_\tau^\gamma$ is only the larger of the
$|A|(|C|)$.

\bigskip

\noindent {\large \bf IV. Conclusions}

We have presented closed-form expressions for the correlations
of $O_1$ and $O_2$ parametrized by the real and  imaginary
parts of the edm and wdff and for their standard deviations due
to standard model interactions. We have tabulated for unit
values of these parameters (in units of $e/\sqrt{s}$) the values
of the correlation and standard deviations for a variety of
energies at whichthe $\tau$CF is expected to operate and have
computed the 1 s.d. sensitivities for a modest sample of $10^7$
$\tau^+\tau^-$ pairs. A polarization asymmetry we define is a
useful tool to improve this sensitivity. We have described a
technique to implement a model independent
analysis by varying the polarization which does not require
us to neglect the contributions of a possible wdff that is
justified when the edm and wdff are of comparable magnitudes.
For $e^+$ and $e^-$
longitudinal beam polarizations of 42\% achievable at the
$\tau$cF the sensitivities can be as excellent as (few)$\cdot
10^{-19} e$ cm for the real part and (few)$\cdot
10^{-16} e$ cm for the imaginary part. We demonstrate that the
absence of an axial vector coupling of the electron to the
photon is not necessarily a detriment to the use of polarization
in probing $CP$ violation. An  improvement by at least an order
of magnitude over the sensitivity for
the real part of the edm in the unpolarized case (Table 2 of
ref. [10]) is noted.

\bigskip

\noindent {\bf  Acknowledgements:} B.A. thanks the Swiss National
Science Foundation for support during the course of this work. A
useful conversation with J. Kirkby is gratefully acknowledged.

\newpage
\noindent{\large \bf References}
\vskip .5cm
\noindent [1] See for instance, C. Jarlskog (ed.), {\it CP Violation},
World Scientific, Singapore, 1989.

\noindent [2] For a review, see S.M. Barr and W. Marciano in [1].

\noindent [3] Review of Particle Properties 94,
L. Montanet {\it et al.}, Phys. Rev. D 50,
1173 (1994).

\noindent [4] For a recent comprehensive
discussion on $\tau$ properties including
CP-violating form factors, see for instance, A. Pich, CERN re\-ports
CERN-TH.7065/93 and CERN-TH.7066/93.

\noindent [5] B. Ananthanarayan and S.D. Rindani, Phys. Rev.
Lett. {\bf 73}, 1215 (1994), Phys. Rev. D {\bf 50}, 4447 (1994).

\noindent [6] For $CP$-odd correlations for the purpose of
studying $CP$ violation, see J.F. Donoghue and G. Valencia,
Phys. Rev. Lett. {\bf 58}, 451
(1987); M. Nowakowski and A. Pilaftsis, Mod. Phys. Lett. A {\bf
4}, 829 (1989), Z. Phys. C {\bf 42}, 449
(1989);   W.
Bernreuther {\it et al.}, Z. Phys. C {\bf 43}, 117 (1989); W.
Bernreuther and O. Nachtmann, Phys. Lett. B {\bf 268}, 424
(1991); M.B. Gavela {\it et al.}, Phys. Rev. D {\bf 39}, 1870
(1989). In the $\tau$ system we consider $CP$-odd correlations
among final state momenta since they are correlated, in turn,
to the spins of the parent $\tau^+$ and $\tau^-$ as discussed
prior to their discovery by  Y.S. Tsai [Phys. Rev. {\bf
D4}, 2821 (1971); {\bf 13},
771 (1976) (E)]. See also  S. Kawasaki, T. Shirafuji and
Y.S. Tsai, Prog. Theo. Phys. {\bf 49}, 1656 (1973).

\noindent [7] F. Hoogeveen and L. Stodolsky, Phys. Lett. B {\bf 212},
505 (1988); S. Goozovat and C.A. Nelson, Phys. Lett. B {\bf
267}, 128 (1991); G. Couture, Phys. Lett. B {\bf 272}, 404
(1991); W. Bernreuther and O. Nachtmann, Phys. Rev. Lett. {\bf
63}, 2787 (1989).

\noindent [8] W. Bernreuther and O. Nachtmann in [7];  W.
Bernreuther {\it et al.}, Z. Phys. C {\bf 52}, 567 (1991).

\noindent [9] OPAL Collaboration, P.D. Acton {\it et al.}, Phys.
Lett. B {\bf
281}, 405 (1992); ALEPH Collaboration, D. Buskulic {\it et al.}, Phys.
Lett. B {\bf 297}, 459 (1992).  Preliminary results show that these
constraints may be even further strengthened to the bounds on the
real and imaginary parts of the wdff of $6.4\cdot 10^{-18}$ and
$4.5\cdot 10^{-17} e$ cm respectively [A. Stahl, Talk given
at the Third Workshop on Tau Lepton Physics, Montreux, Switzerland,
September 1994].

\noindent [10] W. Bernreuther, O. Nachtmann and P. Overmann,
Phys. Rev. D {\bf 48}, 78 (1993).

\noindent [11] See J. Kirkby, CERN reports CERN-PPE/92-30 (1992)
and CERN-PPE/94-37 (1994); N. Qi, Talk given at the Third Workshop
on Tau Lepton Physics, Montreux, Switzerland, September 1994.

\noindent [12] A. Zholents, CERN report CERN SL 92-27 (AP) (1992).

\noindent [13] The case of $t\overline{t}$ production has been
considered in F.~Cuypers and S.D.~Rindani, Munich preprint
MPI-PhT/94-54 (1994), hep-ph/9409243, to appear in Phys. Lett. B.

\noindent [14] It has been shown by F. Cuypers and S. D. Rindani
[13], and P. Poulose and S.D. Rindani [PRL Ahmedabad preprint
PRL-TH-94/31 (1994), hep-ph/9410357]
in the context of the $t\bar{t}$ system
that a polarization experiment such as the one considered here
possesses the power to resolve the edm and wdf contributions by
varying the polarization.  A similar procedure can be adopted at
the $\tau$cF and we defer such a discussion to a later section.

\noindent [15] And yet, it is worthy of note that these
correlations and their standard deviations yield an important
normalization tool for Monte Carlo simulations  that have to be constructed in
order to compute three (and larger) body final states and tensor
correlations, making them {\em of more than academic
interest} even in the event no polarization experiment is performed.

\noindent [16] An alternative proposal is to polarize the electron
beam alone [J. Kirkby, private communication] with the possibility of
a degree of polarization of nearly $100\%$, as for example at
SLC where an $80\%$ polarization has now been achieved [B. Schumm,
Talk given at Third Workshop on Tau Lepton Physics, Montreux, Switzerland,
September 1994].  However, a hard collinear photon bremmstrahlung
process would induce a CP-even background to the correlation $\langle
O_2 \rangle$
at $O(\alpha)$.
Such a background would have to computed and subtracted.  It is possible
to do this in the formalism applicable
to helicity-flip bremmstrahlung following B. Falk and L. M. Sehgal,
Phys. Lett. B 325, 509 (1994).  Note that at SLC the cross-section
of this process is
suppressed by a large factor compared to the production cross-section
(in fact vanishing in the narrow width approximation)
but is not necessarily so at the $\tau$cF energies.

\noindent [17] After the completion of this work we received a paper
by Y-S. Tsai, SLAC report, SLAC-PUB-6685, hep-ph/9410265,
wherein  the use of longitudinally polarized beams to probe CP violation
in $\tau$-decays is discussed.

\newpage
\noindent {\large \bf Table Captions}
\vskip .25cm

\noindent 1. (a) $c_{AB}$, standard deviation and $\vert\delta
{\rm Re}\,
d_{\tau}^{\gamma}\vert$ computed for $10^7$ $\tau^+\tau^-$ pairs
for $\pi\pi$ channel and $\sqrt{s}=3.67$ GeV for operator $O_1$
for different $P$.\\
\noindent ~~(b) Same as above for $\pi\rho$ channel.\\
\noindent ~~(c) Same as above for $\rho\rho$ channel.

\noindent 2. (a) $c_{AB}$, standard deviation and $\vert\delta {\rm Im}\,
d_{\tau}^{\gamma}\vert$ computed for $10^7$ $\tau^+\tau^-$ pairs
for $\pi\pi$ channel and $\sqrt{s}=3.67$ GeV for operator $O_2$
for different $P$.\\
\noindent ~~(b) Same as above for $\pi\rho$ channel.\\
\noindent ~~(c) Same as above for $\rho\rho$ channel.

\noindent 3. Same as (1) for $\sqrt{s}=4.25$ GeV.

\noindent 4. Same as (2) for $\sqrt{s}=4.25$ GeV.

\noindent 5. Same as (1) for $\sqrt{s}=10.58$ GeV.

\noindent 6. Same as (2) for $\sqrt{s}=10.58$ GeV.

\noindent 7. (a) $c_{AB}$, standard deviation and $\vert\delta {\rm Re}\,
d_{\tau}^{\gamma}\vert$ computed for $\int {\cal L} (P) dt =
\int {\cal L} (-P) dt$ and $\sum_{i,j} N_{ij}=2\cdot 10^7$
$\tau^+\tau^-$ pairs  from $O_1$ for $\pi\pi$, $\pi\rho$ and
$\rho\rho$ channels for $\sqrt{s}=3.67$ GeV from polarization
asymmetrized distribution.\\
{}~~(b) $c_{AB}$, standard deviation and $\vert\delta {\rm Im}\,
d_{\tau}^{\gamma}\vert$ computed for $\int {\cal L} (P) dt =
\int {\cal L} (-P) dt$ and $\sum_{i,j} N_{ij}=2\cdot 10^7$
$\tau^+\tau^-$ pairs  from $O_2$ for $\pi\pi$, $\pi\rho$ and
$\rho\rho$ channels for $\sqrt{s}=3.67$ GeV from polarization
asymmetrized distribution.

\noindent 8. Same as (7) for $\sqrt{s}=4.25$ GeV.

\noindent 9. Same as (7) for $\sqrt{s}=10.58$ GeV.

\noindent 10. The effective
polarization asymmetry
$ |\sum_{i,j} N_{ij} r_{ij}|/\left(\sum_{i,j} N_{ij} \right)$
for $\sqrt{s}=3.67$, 4.25 and 10.58 GeV.

\noindent 11. (a) The quantities $a,\ b,\ c $ and $d$ in $e$ cm defining the
lines of sensitivity for different polarization
with $N_{\tau^+\tau^-}=10^7$ for $\sqrt{s}=3.67$ GeV.\\
{}~~(b) As above for $\sqrt{s}=4.25$ GeV.\\
{}~~(c) As above for $\sqrt{s}=10.58$ GeV.

\noindent 12. The quantities
$a, \ b, \ c$ and $d$ for the asymmetrized
distributions with $P=1$, $\int {\cal L}(P) dt=
\int {\cal L}(-P) dt$ and $\sum_{ij}N_{ij}
=2\cdot 10^7$ for the three different center of mass
energies.

\noindent 13. (a) $A$ and $B$, and $C$ and $D$, defining the
parallelograms in the Re$d_\tau^\gamma$ -- Re$d_\tau^Z$ and
Im$d_\tau^\gamma$ -- Im$d_\tau^Z$ planes respectively for various
values of $P_{max}$ for $a,\ b,\ c$ and $d$ of Table 11
as described in the text at $\sqrt{s}=3.67\ {\rm GeV}$.\\
{}~~(b) As above for $\sqrt{s}=4.25\ {\rm GeV}$.\\
{}~~(c) As above for $\sqrt{s}=10.58\ {\rm GeV}$.

\newpage
\begin{center}
\begin{tabular}{||c|r|c|r||}
\hline
$P$&$c_{AB}$ GeV$^2$&$\sqrt{\langle
O_1^2\rangle}$ GeV$^2$&$\vert \delta\,{\rm
Re}\,d_{\tau}^{\gamma}\vert$ $e$ cm\\
\hline
0.00 & $-3.12\times 10^{-6}$ & 0.399 & $1.88\times 10^{-12}$\\
$-0.62$ & $-1.36\times 10^{-2}$ & 0.399 & $4.32\times 10^{-16}$\\
+0.62 & $1.35\times 10^{-2}$ & 0.399 & $4.32\times 10^{-16}$\\
$-0.71$ & $-1.55\times 10^{-2}$ & 0.399 & $3.77\times 10^{-16}$\\
+0.71 & $1.55\times 10^{-2}$ & 0.399 & $3.77\times 10^{-16}$\\
$-1.00$ & $-2.19\times 10^{-2}$ & 0.399 & $2.68\times 10^{-16}$\\
+1.00 & $2.19\times 10^{-2}$ & 0.399 & $2.68\times 10^{-16}$\\
\hline
\multicolumn{4}{c}{}\\
\multicolumn{4}{c}{(a)}\\
\multicolumn{4}{c}{}\\
\hline
$P$&$c_{AB}$ GeV$^2$&$\sqrt{\langle
O_1^2\rangle}$ GeV$^2$&$\vert \delta\,{\rm
Re}\,d_{\tau}^{\gamma}\vert$ $e$ cm\\
\hline
0.00 & $-7.72\times 10^{-7}$ & 0.336 & $4.64\times 10^{-12}$\\
$-0.62$ & $-1.05\times 10^{-2}$ & 0.336 & $3.39\times 10^{-16}$\\
+0.62 & $1.05\times 10^{-2}$ & 0.336 & $3.39\times 10^{-16}$\\
$-0.71$ & $-1.21\times 10^{-2}$ & 0.336 & $2.96\times 10^{-16}$\\
+0.71 & $1.21\times 10^{-2}$ & 0.336 & $2.96\times 10^{-16}$\\
$-1.00$ & $-1.70\times 10^{-2}$ & 0.336 & $2.10\times 10^{-16}$\\
+1.00 & $1.70\times 10^{-2}$ & 0.336 & $2.10\times 10^{-16}$\\
\hline
\multicolumn{4}{c}{}\\
\multicolumn{4}{c}{(b)}\\
\multicolumn{4}{c}{}\\
\hline
$P$&$c_{AB}$ GeV$^2$&$\sqrt{\langle
O_1^2\rangle}$ GeV$^2$&$\vert \delta\,{\rm
Re}\,d_{\tau}^{\gamma}\vert$ $e$ cm\\
\hline
0.00 & $-1.39\times 10^{-7}$ & 0.282 & $1.57\times 10^{-11}$\\
$-0.62$ & $-6.02\times 10^{-3}$ & 0.282 & $3.63\times 10^{-16}$\\
+0.62 & $6.02\times 10^{-3}$ & 0.282 & $3.63\times 10^{-16}$\\
$-0.71$ & $-6.89\times 10^{-3}$ & 0.282 & $3.17\times 10^{-16}$\\
+0.71 & $6.89\times 10^{-3}$ & 0.282 & $3.17\times 10^{-16}$\\
$-1.00$ & $-9.70\times 10^{-3}$ & 0.282 & $2.25\times 10^{-16}$\\
+1.00 & $9.70\times 10^{-3}$ & 0.282 & $2.25\times 10^{-16}$\\
\hline
\multicolumn{4}{c}{}\\
\multicolumn{4}{c}{(c)}\\
\multicolumn{4}{c}{}\\
\end{tabular}
\vskip .2cm
{Table 1}
\end{center}

\newpage
\begin{center}
\begin{tabular}{||c|r|c|r||}
\hline
$P$&$c_{AB}$ GeV&$\sqrt{\langle
O_2^2\rangle}$ GeV&$\vert \delta\,{\rm
Im}\,d_{\tau}^{\gamma}\vert$ $e$ cm\\
\hline
0.00 & $1.36\times 10^{-5}$ & 0.596 & $6.44\times 10^{-13}$\\
$-0.62$ & $1.42\times 10^{-5}$ & 0.596 & $6.14\times 10^{-13}$\\
+0.62 & $1.29\times 10^{-5}$ & 0.596 & $6.78\times 10^{-13}$\\
$-0.71$ & $1.43\times 10^{-5}$ & 0.596 & $6.10\times 10^{-13}$\\
+0.71 & $1.28\times 10^{-5}$ & 0.596 & $6.83\times 10^{-13}$\\
$-1.00$ & $1.46\times 10^{-5}$ & 0.596 & $5.97\times 10^{-13}$\\
+1.00 & $1.25\times 10^{-5}$ & 0.596 & $7.01\times 10^{-13}$\\
\hline
\multicolumn{4}{c}{}\\
\multicolumn{4}{c}{(a)}\\
\multicolumn{4}{c}{}\\
\hline
$P$&$c_{AB}$ GeV&$\sqrt{\langle
O_2^2\rangle}$ GeV&$\vert \delta\,{\rm
Im}\,d_{\tau}^{\gamma}\vert$ $e$ cm\\
\hline
0.00 & $9.33\times 10^{-6}$ & 0.615 & $7.02\times 10^{-13}$\\
$-0.62$ & $9.79\times 10^{-6}$ & 0.615 & $6.69\times 10^{-13}$\\
+0.62 & $8.86\times 10^{-6}$ & 0.615 & $7.39\times 10^{-13}$\\
$-0.71$ & $9.86\times 10^{-6}$ & 0.615 & $6.64\times 10^{-13}$\\
+0.71 & $8.88\times 10^{-6}$ & 0.615 & $7.44\times 10^{-13}$\\
$-1.00$ & $1.01\times 10^{-5}$ & 0.615 & $6.50\times 10^{-13}$\\
+1.00 & $8.58\times 10^{-6}$ & 0.615 & $7.63\times 10^{-13}$\\
\hline
\multicolumn{4}{c}{}\\
\multicolumn{4}{c}{(b)}\\
\multicolumn{4}{c}{}\\
\hline
$P$&$c_{AB}$ GeV&$\sqrt{\langle
O_2^2\rangle}$ GeV&$\vert \delta\,{\rm
Im}\,d_{\tau}^{\gamma}\vert$ $e$ cm\\
\hline
 0.00 & $5.10\times 10^{-6}$ & 0.575 & $8.72\times 10^{-13}$\\
$-0.62$ & $5.35\times 10^{-6}$ & 0.575 & $8.31\times 10^{-13}$\\
+0.62 & $4.85\times 10^{-6}$ & 0.575 & $9.18\times 10^{-13}$\\
$-0.71$ & $5.39\times 10^{-6}$ & 0.575 & $8.25\times 10^{-13}$\\
+0.71 & $4.81\times 10^{-6}$ & 0.575 & $9.25\times 10^{-13}$\\
$-1.00$ & $5.51\times 10^{-6}$ & 0.575 & $8.08\times 10^{-13}$\\
+1.00 & $4.69\times 10^{-6}$ & 0.575 & $9.48\times 10^{-13}$\\
\hline
\multicolumn{4}{c}{}\\
\multicolumn{4}{c}{(c)}\\
\multicolumn{4}{c}{}\\
\end{tabular}
\vskip .2cm
{Table 2}
\end{center}

\newpage
\begin{center}
\begin{tabular}{||c|r|c|r||}
\hline
$P$&$c_{AB}$ GeV$^2$&$\sqrt{\langle
O_1^2\rangle}$ GeV$^2$&$\vert \delta\,{\rm
Re}\,d_{\tau}^{\gamma}\vert$ $e$ cm\\
\hline
0.00 & $-2.67\times 10^{-5}$ & 0.543 & $2.58\times 10^{-13}$\\
$-0.62$ & $-8.64\times 10^{-2}$ & 0.543 & $7.96\times 10^{-17}$\\
+0.62 & $8.63\times 10^{-2}$ & 0.543 & $7.97\times 10^{-17}$\\
$-0.71$ & $-9.89\times 10^{-2}$ & 0.543 & $6.95\times 10^{-17}$\\
+0.71 & $9.88\times 10^{-2}$ & 0.543 & $6.96\times 10^{-17}$\\
$-1.00$ & $-1.39\times 10^{-1}$ & 0.543 & $4.94\times 10^{-17}$\\
+1.00 & $1.39\times 10^{-1}$ & 0.543 & $4.94\times 10^{-17}$\\
\hline
\multicolumn{4}{c}{}\\
\multicolumn{4}{c}{(a)}\\
\multicolumn{4}{c}{}\\
\hline
$P$&$c_{AB}$ GeV$^2$&$\sqrt{\langle
O_1^2\rangle}$ GeV$^2$&$\vert \delta\,{\rm
Re}\,d_{\tau}^{\gamma}\vert$ $e$ cm\\
\hline
0.00 & $-6.59\times 10^{-6}$ & 0.488 & $6.80\times 10^{-13}$\\
$-0.62$ & $-6.72\times 10^{-2}$ & 0.488 & $6.68\times 10^{-17}$\\
+0.62 & $6.72\times 10^{-2}$ & 0.488 & $6.68\times 10^{-17}$\\
$-0.71$ & $-7.69\times 10^{-2}$ & 0.488 & $5.83\times 10^{-17}$\\
+0.71 & $7.69\times 10^{-2}$ & 0.488 & $5.83\times 10^{-17}$\\
$-1.00$ & $-1.08\times 10^{-1}$ & 0.488 & $4.14\times 10^{-17}$\\
+1.00 & $1.08\times 10^{-1}$ & 0.488 & $4.14\times 10^{-17}$\\
\hline
\multicolumn{4}{c}{}\\
\multicolumn{4}{c}{(b)}\\
\multicolumn{4}{c}{}\\
\hline
$P$&$c_{AB}$ GeV$^2$&$\sqrt{\langle
O_1^2\rangle}$ GeV$^2$&$\vert \delta\,{\rm
Re}\,d_{\tau}^{\gamma}\vert$ $e$ cm\\
\hline
0.00 & $-1.19\times 10^{-6}$ & 0.432 & $2.43\times 10^{-12}$\\
$-0.62$ & $-3.83\times 10^{-2}$ & 0.432 & $7.53\times 10^{-17}$\\
+0.62 & $3.83\times 10^{-2}$ & 0.432 & $7.53\times 10^{-17}$\\
$-0.71$ & $-4.39\times 10^{-2}$ & 0.432 & $6.57\times 10^{-17}$\\
+0.71 & $4.39\times 10^{-2}$ & 0.432 & $6.57\times 10^{-17}$\\
$-1.00$ & $-6.18\times 10^{-2}$ & 0.432 & $4.67\times 10^{-17}$\\
+1.00 & $6.18\times 10^{-2}$ & 0.432 & $4.67\times 10^{-17}$\\
\hline
\multicolumn{4}{c}{}\\
\multicolumn{4}{c}{(c)}\\
\multicolumn{4}{c}{}\\
\end{tabular}
\vskip .2cm
{Table 3}
\end{center}

\newpage
\begin{center}
\begin{tabular}{||c|r|c|r||}
\hline
$P$&$c_{AB}$ GeV&$\sqrt{\langle
O_2^2\rangle}$ GeV&$\vert \delta\,{\rm
Im}\,d_{\tau}^{\gamma}\vert$ $e$ cm\\
\hline
0.00 & $1.00\times 10^{-4}$ & 0.632 & $8.00\times 10^{-14}$\\
$-0.62$ & $1.05\times 10^{-4}$ & 0.632 & $7.62\times 10^{-14}$\\
+0.62 & $9.51\times 10^{-5}$ & 0.632 & $8.41\times 10^{-14}$\\
$-0.71$ & $1.06\times 10^{-4}$ & 0.632 & $7.57\times 10^{-14}$\\
+0.71 & $9.44\times 10^{-5}$ & 0.632 & $8.48\times 10^{-14}$\\
$-1.00$ & $1.08\times 10^{-4}$ & 0.632 & $7.40\times 10^{-14}$\\
+1.00 & $9.20\times 10^{-5}$ & 0.632 & $8.69\times 10^{-14}$\\
\hline
\multicolumn{4}{c}{}\\
\multicolumn{4}{c}{(a)}\\
\multicolumn{4}{c}{}\\
\hline
$P$&$c_{AB}$ GeV&$\sqrt{\langle
O_2^2\rangle}$ GeV&$\vert \delta\,{\rm
Im}\,d_{\tau}^{\gamma}\vert$ $e$ cm\\
\hline
0.00 & $6.89\times 10^{-5}$ & 0.655 & $8.74\times 10^{-14}$\\
$-0.62$ & $7.23\times 10^{-5}$ & 0.655 & $8.33\times 10^{-14}$\\
+0.62 & $6.54\times 10^{-5}$ & 0.655 & $9.20\times 10^{-14}$\\
$-0.71$ & $7.28\times 10^{-5}$ & 0.655 & $8.27\times 10^{-14}$\\
+0.71 & $6.49\times 10^{-5}$ & 0.655 & $9.27\times 10^{-14}$\\
$-1.00$ & $7.44\times 10^{-5}$ & 0.655 & $8.09\times 10^{-14}$\\
+1.00 & $6.33\times 10^{-5}$ & 0.655 & $9.50\times 10^{-14}$\\
\hline
\multicolumn{4}{c}{}\\
\multicolumn{4}{c}{(b)}\\
\multicolumn{4}{c}{}\\
\hline
$P$&$c_{AB}$ GeV&$\sqrt{\langle
O_2^2\rangle}$ GeV&$\vert \delta\,{\rm
Im}\,d_{\tau}^{\gamma}\vert$ $e$ cm\\
\hline
 0.00 & $3.76\times 10^{-5}$ & 0.610 & $1.08\times 10^{-13}$\\
$-0.62$ & $3.95\times 10^{-5}$ & 0.610 & $1.03\times 10^{-13}$\\
+0.62 & $3.58\times 10^{-5}$ & 0.610 & $1.14\times 10^{-13}$\\
$-0.71$ & $3.98\times 10^{-5}$ & 0.610 & $1.02\times 10^{-13}$\\
+0.71 & $3.55\times 10^{-5}$ & 0.610 & $1.15\times 10^{-13}$\\
$-1.00$ & $4.06\times 10^{-5}$ & 0.610 & $1.00\times 10^{-13}$\\
+1.00 & $3.46\times 10^{-5}$ & 0.610 & $1.18\times 10^{-13}$\\
\hline
\multicolumn{4}{c}{}\\
\multicolumn{4}{c}{(c)}\\
\multicolumn{4}{c}{}\\
\end{tabular}
\vskip .2cm
{Table 4}
\end{center}

\newpage
\begin{center}
\begin{tabular}{||c|r|c|r||}
\hline
$P$&$c_{AB}$ GeV$^2$&$\sqrt{\langle
O_1^2\rangle}$ GeV$^2$&$\vert \delta\,{\rm
Re}\,d_{\tau}^{\gamma}\vert$ $e$ cm\\
\hline
0.00 & $-1.58\times 10^{-5}$ & 1.78 & $5.71\times 10^{-15}$\\
$-0.62$ & $-8.19\times 10^{-1}$ & 1.78 & $1.10\times 10^{-17}$\\
+0.62 & $8.15\times 10^{-1}$ & 1.78 & $1.11\times 10^{-17}$\\
$-0.71$ & $-9.37\times 10^{-1}$ & 1.78 & $9.65\times 10^{-18}$\\
+0.71 & $9.33\times 10^{-1}$ & 1.78 & $9.67\times 10^{-18}$\\
$-1.00$ &\multicolumn{1}{c|}{ $-1.32$} & 1.78 & $6.86\times 10^{-18}$\\
+1.00 &  \multicolumn{1}{c|}{$1.31$}&1.78 & $6.86\times 10^{-18}$\\
\hline
\multicolumn{4}{c}{}\\
\multicolumn{4}{c}{(a)}\\
\multicolumn{4}{c}{}\\
\hline
$P$&$c_{AB}$ GeV$^2$&$\sqrt{\langle
O_1^2\rangle}$ GeV$^2$&$\vert \delta\,{\rm
Re}\,d_{\tau}^{\gamma}\vert$ $e$ cm\\
\hline
0.00 & $-3.91\times 10^{-4}$ & 1.66 & $1.57\times 10^{-14}$\\
$-0.62$ & $-6.36\times 10^{-1}$ & 1.66 & $9.63\times 10^{-18}$\\
+0.62 & $6.35\times 10^{-1}$ & 1.66 & $9.64\times 10^{-18}$\\
$-0.71$ & $-7.29\times 10^{-1}$ & 1.66 & $8.41\times 10^{-18}$\\
+0.71 & $7.27\times 10^{-1}$ & 1.66 & $8.42\times 10^{-18}$\\
$-1.00$ & \multicolumn{1}{c|}{$- 1.03$}& 1.66 & $5.98\times 10^{-18}$\\
+1.00  &\multicolumn{1}{c|}{$ 1.02$}& 1.66 & $5.98\times 10^{-18}$\\
\hline
\multicolumn{4}{c}{}\\
\multicolumn{4}{c}{(b)}\\
\multicolumn{4}{c}{}\\
\hline
$P$&$c_{AB}$ GeV$^2$&$\sqrt{\langle
O_1^2\rangle}$ GeV$^2$&$\vert \delta\,{\rm
Re}\,d_{\tau}^{\gamma}\vert$ $e$ cm\\
\hline
0.00 & $-7.03\times 10^{-5}$ & 1.51 & $5.76\times 10^{-14}$\\
$-0.62$ & $-3.63\times 10^{-1}$ & 1.51 & $1.12\times 10^{-17}$\\
+0.62 & $3.62\times 10^{-1}$ & 1.51 & $1.12\times 10^{-17}$\\
$-0.71$ & $-4.15\times 10^{-1}$ & 1.51 & $9.77\times 10^{-18}$\\
+0.71 & $4.15\times 10^{-1}$ & 1.51 & $9.77\times 10^{-18}$\\
$-1.00$ & $-5.85\times 10^{-1}$ & 1.51 & $6.93\times 10^{-18}$\\
+1.00 & $5.85\times 10^{-1}$ & 1.51 & $6.94\times 10^{-18}$\\
\hline
\multicolumn{4}{c}{}\\
\multicolumn{4}{c}{(c)}\\
\multicolumn{4}{c}{}\\
\end{tabular}
\vskip .2cm
{Table 5}
\end{center}

\newpage
\begin{center}
\begin{tabular}{||c|r|c|r||}
\hline
$P$&$c_{AB}$ GeV&$\sqrt{\langle
O_2^2\rangle}$ GeV&$\vert \delta\,{\rm
Im}\,d_{\tau}^{\gamma}\vert$ $e$ cm\\
\hline
0.00 & $2.83\times 10^{-3}$ & 1.19 & $2.53\times 10^{-15}$\\
$-0.62$ & $2.50\times 10^{-3}$ & 1.19 & $2.41\times 10^{-15}$\\
+0.62 & $2.26\times 10^{-3}$ & 1.19 & $2.67\times 10^{-15}$\\
$-0.71$ & $2.52\times 10^{-3}$ & 1.19 & $2.40\times 10^{-15}$\\
+0.71 & $2.25\times 10^{-3}$ & 1.19 & $2.69\times 10^{-15}$\\
$-1.00$ & $2.58\times 10^{-3}$ & 1.19 & $2.34\times 10^{-15}$\\
+1.00 & $2.19\times 10^{-3}$ & 1.19 & $2.75\times 10^{-15}$\\
\hline
\multicolumn{4}{c}{}\\
\multicolumn{4}{c}{(a)}\\
\multicolumn{4}{c}{}\\
\hline
$P$&$c_{AB}$ GeV&$\sqrt{\langle
O_2^2\rangle}$ GeV&$\vert \delta\,{\rm
Im}\,d_{\tau}^{\gamma}\vert$ $e$ cm\\
\hline
0.00 & $1.64\times 10^{-3}$ & 1.26 & $2.83\times 10^{-15}$\\
$-0.62$ & $1.72\times 10^{-3}$ & 1.26 & $2.69\times 10^{-15}$\\
+0.62 & $1.56\times 10^{-3}$ & 1.26 & $2.98\times 10^{-15}$\\
$-0.71$ & $1.73\times 10^{-3}$ & 1.26 & $2.67\times 10^{-15}$\\
+0.71 & $1.55\times 10^{-3}$ & 1.26 & $3.00\times 10^{-15}$\\
$-1.00$ & $1.77\times 10^{-3}$ & 1.26 & $2.62\times 10^{-15}$\\
+1.00 & $1.51\times 10^{-3}$ & 1.26 & $3.08\times 10^{-15}$\\
\hline
\multicolumn{4}{c}{}\\
\multicolumn{4}{c}{(b)}\\
\multicolumn{4}{c}{}\\
\hline
$P$&$c_{AB}$ GeV&$\sqrt{\langle
O_2^2\rangle}$ GeV&$\vert \delta\,{\rm
Im}\,d_{\tau}^{\gamma}\vert$ $e$ cm\\
\hline
 0.00 & $8.96\times 10^{-4}$ & 1.15 & $3.43\times 10^{-15}$\\
$-0.62$ & $9.41\times 10^{-4}$ & 1.15 & $3.26\times 10^{-15}$\\
+0.62 & $8.51\times 10^{-4}$ & 1.15 & $3.61\times 10^{-15}$\\
$-0.71$ & $9.48\times 10^{-4}$ & 1.15 & $3.24\times 10^{-15}$\\
+0.71 & $8.45\times 10^{-4}$ & 1.15 & $3.64\times 10^{-15}$\\
$-1.00$ & $9.69\times 10^{-4}$ & 1.15 & $3.17\times 10^{-15}$\\
+1.00 & $8.24\times 10^{-4}$ & 1.15 & $3.73\times 10^{-15}$\\
\hline
\multicolumn{4}{c}{}\\
\multicolumn{4}{c}{(c)}\\
\multicolumn{4}{c}{}\\
\end{tabular}
\vskip .2cm
{Table 6}
\end{center}

\newpage
\begin{center}
\begin{tabular}{||c|r|c|r||}
\hline
&$c_{AB}$ GeV$^2$&$\sqrt{\langle
O_1^2\rangle}$ GeV$^2$&$\vert \delta\,{\rm
Re}\,d_{\tau}^{\gamma}\vert$ $e$ cm\\
\hline
$\pi\pi$ & $2.39\times 10^{2}$ & 0.292 & $1.32\times 10^{-18}$\\
$\pi\rho$& $1.86\times 10^{2}$ & 0.267 & $1.13\times 10^{-18}$\\
$\rho\rho$ & $1.06\times 10^{2}$ & 0.240 & $1.29\times 10^{-18}$\\
\hline
\multicolumn{4}{c}{}\\
\multicolumn{4}{c}{(a)}\\
\multicolumn{4}{c}{}\\
\hline
&$c_{AB}$ GeV&$\sqrt{\langle
O_2^2\rangle}$ GeV$^2$&$\vert \delta\,{\rm
Im}\,d_{\tau}^{\gamma}\vert$ $e$ cm\\
\hline
$\pi\pi$& $1.18\times 10^{-2}$ & 0.596 & $5.45\times 10^{-14}$\\
$\pi\rho$ & $8.15\times 10^{-3}$ & 0.615 & $5.94\times 10^{-14}$\\
$\rho\rho$ & $4.45\times 10^{-3}$ & 0.575 & $7.38\times 10^{-14}$\\
\hline
\multicolumn{4}{c}{}\\
\multicolumn{4}{c}{(b)}\\
\multicolumn{4}{c}{}\\
\end{tabular}
\vskip .2cm
{Table 7}
\end{center}

\bigskip

\begin{center}
\begin{tabular}{||c|r|c|r||}
\hline
&$c_{AB}$ GeV$^2$&$\sqrt{\langle
O_1^2\rangle}$ GeV$^2$&$\vert \delta\,{\rm
Re}\,d_{\tau}^{\gamma}\vert$ $e$ cm\\
\hline
$\pi\pi$ & $1.13\times 10^{3}$ & 0.529 & $3.77\times 10^{-19}$\\
$\pi\rho$& $8.82\times 10^{2}$ & 0.289& $1.90\times 10^{-19}$\\
$\rho\rho$ & $5.03\times 10^{2}$ & 0.127 & $1.08\times 10^{-19}$\\
\hline
\multicolumn{4}{c}{}\\
\multicolumn{4}{c}{(a)}\\
\multicolumn{4}{c}{}\\
\hline
&$c_{AB}$ GeV&$\sqrt{\langle
O_2^2\rangle}$ GeV$^2$&$\vert \delta\,{\rm
Im}\,d_{\tau}^{\gamma}\vert$ $e$ cm\\
\hline
$\pi\pi$& $6.52\times 10^{-2}$ & 0.632 & $7.83\times 10^{-15}$\\
$\pi\rho$ & $4.47\times 10^{-2}$ & 0.657 & $8.59\times 10^{-15}$\\
$\rho\rho$ & $2.45\times 10^{-2}$ & 0.610 & $1.06\times 10^{-14}$\\
\hline
\multicolumn{4}{c}{}\\
\multicolumn{4}{c}{(b)}\\
\multicolumn{4}{c}{}\\
\end{tabular}
\vskip .2cm
{Table 8}
\end{center}

\newpage
\begin{center}
\begin{tabular}{||c|r|c|r||}
\hline
&$c_{AB}$ GeV$^2$&$\sqrt{\langle
O_1^2\rangle}$ GeV$^2$&$\vert \delta\,{\rm
Re}\,d_{\tau}^{\gamma}\vert$ $e$ cm\\
\hline
$\pi\pi$ & $1.72\times 10^{3}$ & 3.46 & $2.61\times 10^{-19}$\\
$\pi\rho$& $1.34\times 10^{3}$ & 2.38 & $1.68\times 10^{-19}$\\
$\rho\rho$ & $7.62\times 10^{2}$ & 1.48 & $1.33\times 10^{-19}$\\
\hline
\multicolumn{4}{c}{}\\
\multicolumn{4}{c}{(a)}\\
\multicolumn{4}{c}{}\\
\hline
&$c_{AB}$ GeV&$\sqrt{\langle
O_2^2\rangle}$ GeV$^2$&$\vert \delta\,{\rm
Im}\,d_{\tau}^{\gamma}\vert$ $e$ cm\\
\hline
$\pi\pi$& $2.49\times 10^{-1}$ & 1.19 & $6.20\times 10^{-16}$\\
$\pi\rho$ & $1.71\times 10^{-1}$ & 1.28 & $7.03\times 10^{-16}$\\
$\rho\rho$ & $9.35\times 10^{-2}$ & 1.15 & $8.39\times 10^{-16}$\\
\hline
\multicolumn{4}{c}{}\\
\multicolumn{4}{c}{(b)}\\
\multicolumn{4}{c}{}\\
\end{tabular}
\vskip .2cm
{Table 9}
\end{center}

\bigskip

\begin{center}
\begin{tabular}{||c|c||}
\hline
$\sqrt{s}$ GeV&$ |\sum N_{ij}r_{ij}| / \sum N_{ij}$ \\
\hline
3.67& $9.2\times 10^{-5}$ \\
4.25& $1.2\times 10^{-4}$\\
10.58& $7.7\times 10^{-4}$\\
\hline
\end{tabular}
\vskip .2cm
{Table 10}
\end{center}

\newpage

\begin{center}
\begin{tabular}{||c|c|c|c|c||}
\hline
$P$ & $a$ & $b$ & $c$ & $d$ \\
\hline
0.00 & $1.73\times 10^{-12}$ & $1.38\times 10^{-13}$ & $4.17\times 10^{-13}$ &
$2.70\times 10^{-9}$ \\
$-0.62$ & $2.15\times 10^{-16}$ & $1.32 \times 10^{-13}$ & $3.97 \times
10^{-13}$ & $5.52 \times 10^{-10}$\\
$+0.62 $& $2.15 \times 10^{-16}$ & $1.45 \times 10^{-13}$ & $4.39 \times
10^{-13}$ & $9.32 \times 10^{-10}$\\
$-0.71$ & $1.88 \times 10^{-16}$ & $1.31 \times 10^{-13}$ & $3.95 \times
10^{-13}$ & $4.94 \times 10^{-10}$\\
$+0.71$& $1.88 \times 10^{-16}$ & $1.47 \times 10^{-13}$ & $4.42 \times
10^{-13}$ & $7.80 \times 10^{-10}$\\
$-1.00$ & $1.33 \times 10^{-16}$ & $1.28 \times 10^{-13}$ & $3.86 \times
10^{-13}$ & $3.71 \times 10^{-10}$\\
$+1.00$ & $1.33 \times 10^{-16}$ & $1.50 \times 10^{-13}$ & $4.53 \times
10^{-13}$ & $5.11 \times 10^{-10}$\\
\hline
\multicolumn{5}{c}{}\\
\multicolumn{5}{c}{(a)}\\
\multicolumn{5}{c}{}\\
\hline
$P$ & $a$ & $b$ & $c$ & $d$ \\
\hline
$0.00$ & $2.39\times 10^{-13}$ & $2.03\times 10^{-14}$ & $5.18\times 10^{-14}$
& $2.50\times 10^{-10}$ \\
$-0.62$ & $4.23\times 10^{-17}$ & $1.93 \times 10^{-14}$ & $4.93 \times
10^{-14}$ & $5.11 \times 10^{-11}$\\
$+0.62$ & $4.23 \times 10^{-17}$ & $2.13 \times 10^{-14}$ & $5.45 \times
10^{-14}$ & $8.63 \times 10^{-11}$\\
$-0.71$ & $3.69 \times 10^{-17}$ & $1.92 \times 10^{-14}$ & $4.90 \times
10^{-14}$ & $4.58 \times 10^{-10}$\\
$+0.71$ & $3.70 \times 10^{-17}$ & $2.15 \times 10^{-14}$ & $5.49 \times
10^{-14}$ & $7.22 \times 10^{-10}$\\
$-1.00$ & $2.62 \times 10^{-17}$ & $1.88 \times 10^{-14}$ & $4.80 \times
10^{-14}$ & $3.43 \times 10^{-11}$\\
$+1.00$ & $2.62 \times 10^{-17}$ & $2.21 \times 10^{-14}$ & $5.63 \times
10^{-14}$ & $4.73 \times 10^{-11}$\\
\hline
\multicolumn{5}{c}{}\\
\multicolumn{5}{c}{(b)}\\
\multicolumn{5}{c}{}\\
\hline
$P$ & $a$ & $b$ & $c$ & $d$ \\
\hline
0.00 & $5.34\times 10^{-15}$ & $4.66\times 10^{-16}$ & $1.65\times 10^{-15}$ &
$1.27\times 10^{-12}$ \\
$-0.62$ & $6.09\times 10^{-18}$ & $4.44 \times 10^{-16}$ & $1.57 \times
10^{-15}$ & $2.60 \times 10^{-13}$\\
$+0.62 $& $6.09 \times 10^{-18}$ & $4.91 \times 10^{-16}$ & $1.74 \times
10^{-15}$ & $4.40 \times 10^{-13}$\\
$-0.71$ & $5.32 \times 10^{-18}$ & $4.41 \times 10^{-16}$ & $1.56\times
10^{-15}$ & $2.33 \times 10^{-13}$\\
$+0.71$& $5.32 \times 10^{-18}$ & $4.94 \times 10^{-16}$ & $1.75 \times
10^{-15}$ & $3.68 \times 10^{-13}$\\
$-1.00$ & $3.78 \times 10^{-18}$ & $4.31 \times 10^{-16}$ & $1.53 \times
10^{-15}$ & $1.75\times 10^{-13}$\\
$+1.00$ & $3.78 \times 10^{-18}$ & $5.07 \times 10^{-16}$ & $1.80 \times
10^{-15}$ & $2.41\times 10^{-13}$\\
\hline
\multicolumn{5}{c}{}\\
\multicolumn{5}{c}{(c)}\\
\multicolumn{5}{c}{}\\
\end{tabular}
\vskip .2cm
{Table 11}
\end{center}

\newpage

\begin{center}
\begin{tabular}{||c|c|c|c|c||}
\hline
$\sqrt{s}\ {\rm GeV}$ & $a$ & $b$ & $c$ & $d$ \\
\hline
$3.67$ & $7.14\times 10^{-19}$ & $9.29\times 10^{-15}$ & $3.52\times 10^{-14} $
& $ 2.91\times 10^{-12} $\\
$4.25$ & $ 9.11 \times 10^{-20}$ & $8.79\times 10^{-16}$ & $5.07 \times
10^{-15}$ & $3.13 \times 10^{-13} $\\
$10.58$ & $9.68\times 10^{-20}$ & $1.49 \times 10^{-16}$ & $4.06 \times
10^{-16} $ & $3.99 \times 10^{-15} $\\
\hline
\end{tabular}
\vskip .2cm
{Table 12}
\end{center}

\begin{center}
\begin{tabular}{||c|c|c|c|c||}
\hline
$P_{max}$ & $A$ & $B$ & $C$ & $D$ \\
\hline
$0.62$ & $-4.34\times 10^{-15}$ & $2.79 \times 10^{-12}$ & $5.17 \times
10^{-13}$ & $-1.66 \times 10^{-10}$\\
$ $& $2.16 \times 10^{-16}$ & $-4.55 \times 10^{-16}  $ & $-2.02 \times
10^{-12}$ & $3.35 \times 10^{-9}$\\
$0.71$ & $-3.31 \times 10^{-15}$ & $2.44 \times 10^{-12}$ & $5.58 \times
10^{-13}$ & $-2.06 \times 10^{-10}$\\
$ $& $1.88 \times 10^{-16}$ & $ -3.25 \times 10^{-16} $ & $-2.50 \times
10^{-12}$ & $3.62 \times 10^{-9}$\\
$1.00$ & $-1.66 \times 10^{-15}$ & $1.73 \times 10^{-12}$ & $8.40 \times
10^{-13}$ & $-4.36 \times 10^{-10}$\\
$ $ & $1.33 \times 10^{-16}$ & $ -1.94 \times 10^{-16} $ & $-5.28 \times
10^{-12}$ & $5.44 \times 10^{-9}$\\
\hline
\multicolumn{5}{c}{}\\
\multicolumn{5}{c}{(a)}\\
\multicolumn{5}{c}{}\\
\hline
$P_{max}$ & $A$ & $B$ & $C$ & $D$ \\
\hline
$0.62$ & $-8.54 \times 10^{-16}$ & $4.10 \times 10^{-13}$ & $6.42 \times
10^{-14}$ & $-1.54 \times 10^{-11}$\\
$ $ & $4.25 \times 10^{-17}$ & $ -6.78 \times 10^{-17} $ & $-2.50 \times
10^{-13}$ & $3.10 \times 10^{-10}$\\
$0.71$ & $-6.51 \times 10^{-16}$ & $3.58 \times 10^{-13}$ & $6.94 \times
10^{-14}$ & $-1.91 \times 10^{-11}$\\
$ $ & $3.70 \times 10^{-17}$ & $ -4.83 \times 10^{-17} $ & $-3.10 \times
10^{-13}$ & $3.35 \times 10^{-10}$\\
$1.00 $ & $-3.28 \times 10^{-16}$ & $2.54 \times 10^{-13}$ & $1.04 \times
10^{-13}$ & $-4.04 \times 10^{-11}$\\
$ $ & $2.63 \times 10^{-17}$ & $  -3.38 \times 10^{-17} $ & $-6.56 \times
10^{-13}$ & $5.04 \times 10^{-10}$\\
\hline
\multicolumn{5}{c}{}\\
\multicolumn{5}{c}{(b)}\\
\multicolumn{5}{c}{}\\
\hline
$P_{max}$ & $A$ & $B$ & $C$ & $D$ \\
\hline
$0.62$ & $-1.24\times 10^{-16}$ & $9.50 \times 10^{-15}$ & $2.05 \times
10^{-15}$ & $-7.91 \times 10^{-14}$\\
$ $& $6.17 \times 10^{-18}$ & $  -5.61 \times 10^{-18} $ & $-7.99 \times
10^{-15}$ & $1.58 \times 10^{-12}$\\
$0.71$ & $-9.43 \times 10^{-17}$ & $8.26 \times 10^{-15}$ & $2.22 \times
10^{-15}$ & $-9.79 \times 10^{-14}$\\
$ $& $5.36 \times 10^{-18}$ & $  -3.57 \times 10^{-18} $ & $-9.89 \times
10^{-15}$ & $1.71 \times 10^{-12}$\\
$1.00$ & $-4.72 \times 10^{-17}$ & $5.83 \times 10^{-15}$ & $3.34 \times
10^{-15}$ & $-2.07\times 10^{-13}$\\
$ $ & $3.78 \times 10^{-18}$ & $  -3.86 \times 10^{-19} $ & $-2.09 \times
10^{-14}$ & $2.56\times 10^{-12}$\\
\hline
\multicolumn{5}{c}{}\\
\multicolumn{5}{c}{(c)}\\
\multicolumn{5}{c}{}\\
\end{tabular}
\vskip .2cm
{Table 13}
\end{center}

\end{document}